\documentclass[a4paper,11pt]{article}

\usepackage[T1]{fontenc}
\usepackage[utf8]{inputenc}
\usepackage[english]{babel}
\usepackage[left=3cm, right=3cm]{geometry}

\usepackage{lmodern}
\usepackage{amsmath, amsfonts, amssymb, bm}
\usepackage{graphicx}
\usepackage{hyperref}
\usepackage{upgreek}
\usepackage[super,sort&compress]{natbib}
\usepackage{authblk}
\usepackage{lineno}
\usepackage{setspace}
\usepackage{xcolor}
\usepackage{caption}
\usepackage{multirow}
\usepackage{todonotes}


\title{Enrichment of rhombohedral stacking by mechanical exfoliation of graphite}

\author[1,2,3]{Kriszti\'{a}n M\'{a}rity}
\author[1,2,4]{Konr\'{a}d Kandrai}
\author[1]{Gergely Dobrik}
\author[1]{Zsolt E. Horv\'{a}th}
\author[1]{Krist\'{o}f N\'{e}meth D\'{a}niel}
\author[1,5]{Gy\"{o}rgy K\'{a}lvin}
\author[1]{Levente Tapaszt\'{o}}
\author[1,2,*]{P\'{e}ter Nemes-Incze}

\affil[1]{Hungarian Research Network, Centre for Energy Research, Institute of Technical Physics and Materials Science, 1121 Budapest, Hungary}
\affil[2]{MTA - HUN-REN EK Lendület ``Momentum'' Topology in Nanomaterials Research Group, 1121 Budapest, Hungary}
\affil[3]{\'Obuda University, 1081 Budapest, Hungary, Doctoral School of Materials Sciences and
Technologies}
\affil[4]{Budapest University of Technology and Economics, 1111 Budapest, Hungary}
\affil[5]{ELTE E\"{o}tv\"{o}s Lor\'{a}nd University, 1117 Budapest, Hungary}
\affil[*]{\small \emph{Corresponding author: nemes.incze.peter@ek.hun-ren.hu}}

\begin{document}

\maketitle
\doublespacing
% \linenumbers
% \section *{Abstract}
\begin{abstract}
Rhombohedral (ABC) graphite hosts a surface-localized flat band that supports correlated and topological electronic phases, but its experimental study is limited by the scarcity of ABC stacking in natural graphite, which is dominated by Bernal (AB) stacking.
Here we show that the routine mechanical exfoliation step itself enriches the rhombohedral content of graphite flakes, and that a simple blade-assisted exfoliation step, which introduces additional shear, amplifies the effect further.
Using large-area Raman 2D-band skewness mapping we measure ABC content at area fractions of 3\% in the pristine source crystal, 16\% after conventional exfoliation, and 26\% after blade-assisted exfoliation for thick flakes.
In thin flakes ($<20$ layers) the per-flake area fraction reaches 75\% in the upper tail of the distribution.
Tracking individual flakes before and after blade-assisted exfoliation shows that wrinkles seed AB-ABC domain walls, and uniaxial strain can move these walls.
Blade-assisted mechanical exfoliation therefore removes one of the bottlenecks to the preparation of ABC-rich graphite samples for studies of correlated and topological phases in rhombohedral graphite.
\end{abstract}

Graphite is a layered van der Waals material composed of stacked graphene sheets.
While the energetically favored stacking sequence in natural graphite is the Bernal arrangement (AB), also referred to as the hexagonal phase, graphite can also adopt the rhombohedral arrangement (ABC stacking)~\cite{Laves1956-uh}, as well as other intermediate stackings~\cite{Roy2025-su}.
% Throughout this work, we use ``AB'' and ``hexagonal'' interchangeably to designate the Bernal phase, and ``ABC'' and ``rhombohedral'' interchangeably for the rhombohedral phase.
The rhombohedral phase has attracted considerable attention because of its surface-localized flat band, providing a platform for correlated and topological electronic phenomena without the need for moir\'e superlattices~\cite{Han2023-eb,Hagymasi2022-hg,Liu2023-gh,Zhou2021-of,Yang2025-ie}.
Two bottlenecks have so far limited the experimental exploration of rhombohedral graphite: the unambiguous identification of the stacking configuration, and the controlled preparation of ABC-rich samples.
The first has been recently addressed by electronic Raman scattering, which provides an unambiguous fingerprint of stacking-fault-free rhombohedral stacking~\cite{Palinkas2024-sm}.
In this work, we tackle the second.
A variety of routes to prepare the rhombohedral phase have been explored, including chemical vapor deposition on copper~\cite{Bouhafs2021-it}, curvature-stabilized epitaxial growth~\cite{Gao2020-bj}, as well as proposals for shear-~\cite{Nery2020-fe} or strain-driven~\cite{Dey2024-xg} transformations in few-layer graphite.
Despite this growing toolbox, mechanical exfoliation remains the go-to method to prepare high-quality samples for basic research.

Hexagonal and rhombohedral domains are typically found within the same flake, separated by lateral domain walls that accommodate the relative in-plane shifts between the two stacking sequences.
The creation and motion of these domain walls are therefore the elementary processes by which one stacking order transforms into the other.
Bending and curvature have been shown to generate shear across all the graphene layers within a flake~\cite{Korhonen2015-le}, and folds can host stacking faults and graphite twin boundaries with hexagonal-to-rhombohedral character~\cite{Rooney2018-px}.
We show that mechanical exfoliation inevitably introduces bending, curvature, strain, and shear, both during peeling of the adhesive tape and during contact with the substrate, resulting in the formation of domain walls and the rhombohedral phase (see Fig.~\ref{fig:intro}a).
Furthermore, this mechanical perturbation can then displace the resulting domain walls and reorganize the local stacking landscape.

The exfoliation step is most often treated as a passive isolation procedure rather than as an active driver of stacking rearrangement.
Concurrent work has begun to address this question~\cite{Holleis2026-qu}.
Here, we demonstrate that mechanical exfoliation itself can actively promote the formation of rhombohedral stacking in graphite.
Using large-area statistical Raman mapping, we show that exfoliation increases the abundance of ABC domains by nearly an order of magnitude compared to pristine graphite, while additional blade-assisted exfoliation further enhances their occurrence.
We show that pristine natural graphite contains only $\sim$3\% rhombohedral material, consistent with the slightly higher stacking energy of the ABC phase relative to the Bernal phase~\cite{Nery2021-dv}.
A modified exfoliation process, which introduces more shear, raises this fraction by nearly an order of magnitude, up to $\sim$26\%.
Beyond this statistical evidence, we provide direct local proof of a deformation-induced stacking transformation in individual flakes.
Furthermore, by applying uniaxial strain to exfoliated samples, we observe a dynamic evolution of stacking domains and relaxation of stacking-fault regions into energetically favorable configurations.
These results establish mechanical deformation as a practical route for controlling stacking order in graphite.

\section*{Results and Discussion}

In single-crystal natural graphite samples, XRD places the rhombohedral content at 0 to 1\%~\cite{Laves1956-uh}, in agreement with our own measurements on the source crystal used here (see Supporting Information, Section S5).
If this composition were preserved through exfoliation, ABC domains in exfoliated flakes would be exceedingly rare.
The high ABC abundance we report below therefore points to the exfoliation step itself as an active source of the rhombohedral phase.

\begin{figure}[htbp]
    \centering
    \includegraphics[width=0.85 \textwidth]{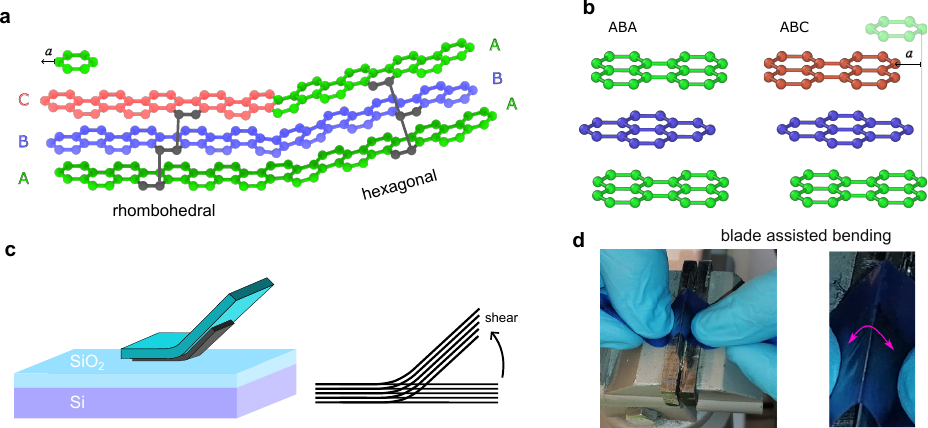}
    \caption{\textbf{Mechanically induced stacking rearrangement in graphite.}
    \textbf{a,} Schematic of the rhombohedral (ABC) and Bernal (AB) stacking configurations in graphite, showing that bending can induce relative lateral displacement between adjacent graphene layers and thereby alter the local stacking order.
    \textbf{b,} Geometric pathway of stacking transformation by lateral sliding: successive in-plane displacement of one graphene layer converts AB stacking into the ABC configuration.
    \textbf{c,} Schematic of micromechanical exfoliation on a Si/SiO$_2$ substrate.
    Shear and bending introduced during exfoliation can promote interlayer sliding and local stacking rearrangement.
    \textbf{d,} Blade-assisted bending applied to the adhesive tape to introduce additional mechanical deformation during sample preparation.}
    \label{fig:intro}
\end{figure}

The hexagonal-to-rhombohedral transformation requires a uniform shear across all the graphene layers of a flake, which bending naturally produces (Fig.~\ref{fig:intro}a-c)~\cite{Korhonen2015-le}.
During exfoliation and release onto the SiO$_2$ substrate, the flakes experience bending and wrinkling that can nucleate domain walls between AB and ABC stacking~\cite{Rooney2018-px}.
The energy cost of moving such domain walls is small~\cite{Halbertal2021-lq}, so they can propagate across a flake under the additional bending and shear delivered by subsequent exfoliation steps.
To amplify these effects, we modified the exfoliation procedure to increase the bending and strain experienced by the flakes.
After the initial peel from the bulk crystal, we redistributed the flakes on the adhesive tape by repeating the exfoliation 30 times.
We then folded the tape against itself to form a tape-graphite-tape sandwich and passed it across a blunt razor blade 30 times to introduce additional bending and shear (Fig.~\ref{fig:intro}d).
The crystals were transferred onto Si/SiO$_2$ wafers for optical and Raman analysis.
We prepared samples in three ways: conventional exfoliation, conventional exfoliation followed by blade-assisted bending, and a variant of the latter in which thermal release tape was used for the final deposition to maximize substrate coverage for statistical Raman mapping (see Supplementary section S6).

We identify hexagonal and rhombohedral domains from the Raman 2D-band line shape using the 2D peak skewness as the primary descriptor.
The skewness is defined as the third standardized moment $\gamma = \mu_3 / \mu_2^{3/2}$ of the Raman intensity values within the 2500-3000 cm$^{-1}$ window, where $\mu_k = N^{-1}\sum_i (I_i - \bar{I})^k$ is the $k$-th central moment of the intensity distribution, $I_i$ is the intensity at the $i$-th spectral point, $\bar{I}$ is its mean over the window, and $N$ is the number of spectral points in the window.
This third standardized moment provides a robust scalar measure of the 2D-band asymmetry, enabling spatial mapping of local stacking variations.
Because mechanical strain can alter the spectral position of the 2D peak, the previously used integrated 2D area-ratio analysis~\cite{Palinkas2024-sm,Yang2019-er} is sensitive to these effects.
Changes in the 2D peak position leave the skewness unchanged, which makes the skewness robust against the strain inhomogeneities found at wrinkles and folds (see Figure S1g).
The Supporting Information provides details of the skewness definition, baseline subtraction, and the comparison with the area-ratio analysis (Section S1, Figure S1).

\begin{figure}[!htbp]
    \centering
    \includegraphics[width=\textwidth]{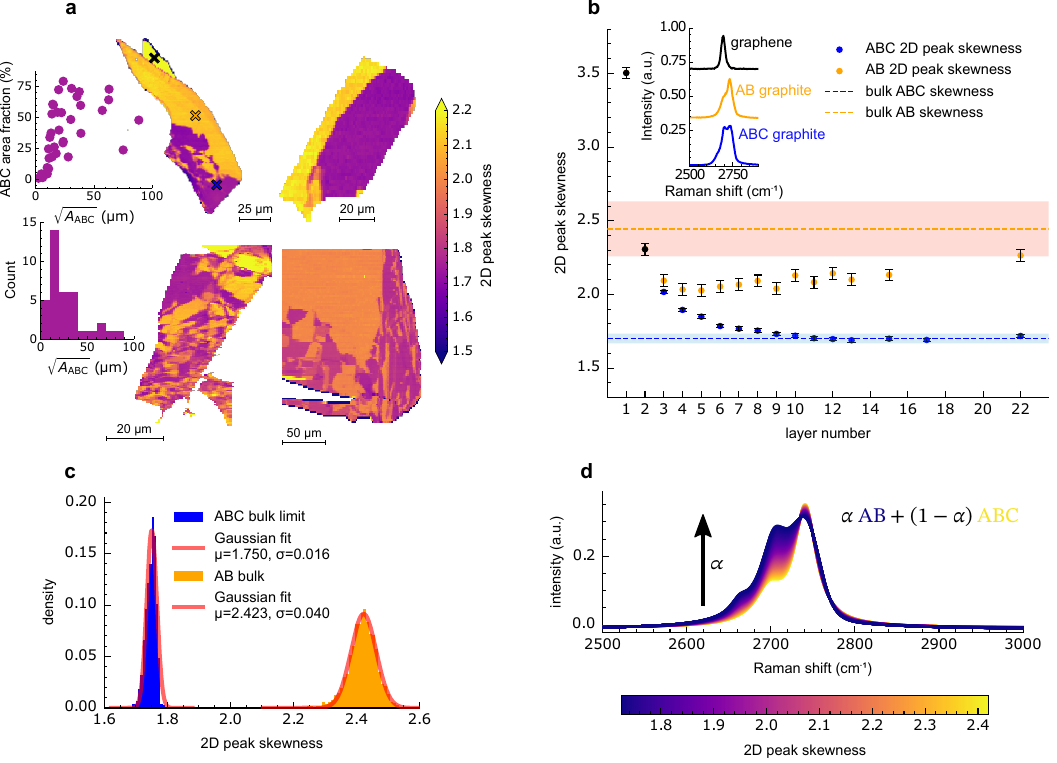}
    \caption{\textbf{2D peak skewness for stacking sequence classification.}
    \textbf{a} Representative 2D peak skewness maps on few-layer flakes showing distinguishable AB- and ABC-stacked domains.
    Insets: distributions of the characteristic size ($\sqrt{\text{ABC area}}$ / flake) of all few-layer flakes examined.
    Purple areas correspond to domains containing ABC stacking sequences.
    \textbf{b} Layer-number dependence of the 2D-band skewness of ABC and AB few-layer flakes, together with reference values for bulk AB and ABC graphite.
    The shaded bands mark the classification windows used in the statistical analysis: the wider band around the bulk AB reference spans $2.25 < \gamma < 2.56$ (spectra below this band are classified as ABC-like; spectra above are excluded from the graphite-covered area), and the narrow band around the bulk ABC reference marks the bulk-ABC distribution shown in panel c.
    Error bars represent the standard deviation of the 2D peak skewness across pixels in a reference Raman map of the corresponding stacking.
    Inset: single Raman spectra at the positions marked by correspondingly colored crosses in a.
    \textbf{c} Histograms of the 2D peak skewness values obtained from Raman maps of homogeneous thick AB and ABC graphite flakes.
    The blue histogram encompasses a sample area of 1500~$\upmu$m$^2$ from a 9-layer homogeneous rhombohedral region, while the orange histogram data result from a 0.26~mm$^2$ area of bulk graphite.
    \textbf{d} Schematic evolution of the Raman 2D band, constructed from Lorentzian fits of bulk AB and ABC 2D peaks.}
    \label{fig:flg_skew}
\end{figure}

Using electronic Raman scattering (ERS) to identify defect-free rhombohedral regions~\cite{Palinkas2024-sm}, we tracked the 2D skewness as a function of layer number from 3 to 22 layers (Fig.~\ref{fig:flg_skew}b).
The skewness of the defect-free rhombohedral phase saturates at 1.7 above 12 layers, so thicker rhombohedral stacks produce an identical 2D line shape.
The hexagonal-phase skewness rises more gradually toward the bulk AB value of 2.42.
Some of the few-layer points at 9, 11, and 13 layers fall below the trend.
Mostly hexagonal polytypes lack sharp ERS features and cannot be distinguished from pure AB by ERS~\cite{McEllistrim2023-op}, so these flakes may contain one or two rhombohedrally stacked layers that depress the skewness.
The 2D peak skewness therefore serves as a fingerprint for the hexagonal or rhombohedral character of a flake, while ERS remains required for unambiguous identification of the exact stacking sequence~\cite{Palinkas2024-sm}.

Examples of few-layer graphene (FLG) flakes prepared by blade-assisted exfoliation are shown in Fig.~\ref{fig:flg_skew}a, where AB- and ABC-stacked domains are resolved by the skewness contrast, with rhombohedral domains showing lower skewness values.
For thin flakes ($<20$ layers) prepared by blade-assisted exfoliation, we analyzed 35 individual flakes with a total mapped area of $0.1$ mm$^2$.
Of this area, $0.039$ mm$^2$ was classified as ABC-stacked graphite (skewness between 1.6 and 1.95, see Fig.~\ref{fig:flg_skew}b), corresponding to a total ABC area fraction of 39\%.
The per-flake ABC area fractions are distributed around 50\% (inset of Fig.~\ref{fig:flg_skew}a), so almost any thin flake produced by blade-assisted exfoliation contains a sizeable rhombohedral region.
We mention that classifying based on the skewness range results in an ABC area estimate that contains not only pure rhombohedral stacking but also intermediate non-hexagonal stacking configurations.

ERS-based statistics over mm$^2$ areas are impractical because of the long measurement times involved, so for the large-area statistical analysis we turn to thicker crystals, with larger flake areas.
We use the bulk AB skewness of 2.42 (Fig.~\ref{fig:flg_skew}c) as the reference value, taken from Raman maps over hexagonal regions of the pristine, unexfoliated graphite source.
Flake areas whose skewness falls below the lower edge of this bulk-AB distribution carry a rhombohedral contribution, and we count them as rhombohedral in the following statistical analysis.
For thick (yellow on 90 nm SiO$_2$) flakes, mixed-stacking regions contain hexagonally stacked layers spanning the crystal thickness, and the 2D line shape can be considered as a linear combination of pure AB and ABC reference spectra (Fig.~\ref{fig:flg_skew}d).
In thin flakes the specific few-layer stacking polytype sets the 2D peak shape rather than a superposition of bulk references, but ABC and AB stacks remain well separated in 2D skewness across the 3--22 layer range (Fig.~\ref{fig:flg_skew}b), so skewness makes possible the identification of stackings that contain ABC sequences in both regimes.
We make no claim about the specific stacking sequence in either case.
For thick flakes, we classify spectra with $1.6 < \gamma < 2.25$ as ABC-like, and define the full graphite-covered area as spectra with $1.6 < \gamma < 2.56$.
The lower bound 1.6 lies at the lower edge of the bulk-ABC distribution (Fig.~\ref{fig:flg_skew}b), set below the asymptotic ABC saturation value of 1.7 so that the full width of the bulk-ABC skewness distribution (blue in Fig.~\ref{fig:flg_skew}c) is included.
The upper bound 2.25 separates the AB-tail of the distribution from the ABC-like values.
The upper bound 2.56 marks where the bulk-AB distribution falls to background (for more details, see Supporting Information, Section S2).

\begin{figure}[htbp]
    \centering
    \includegraphics[width=\textwidth]{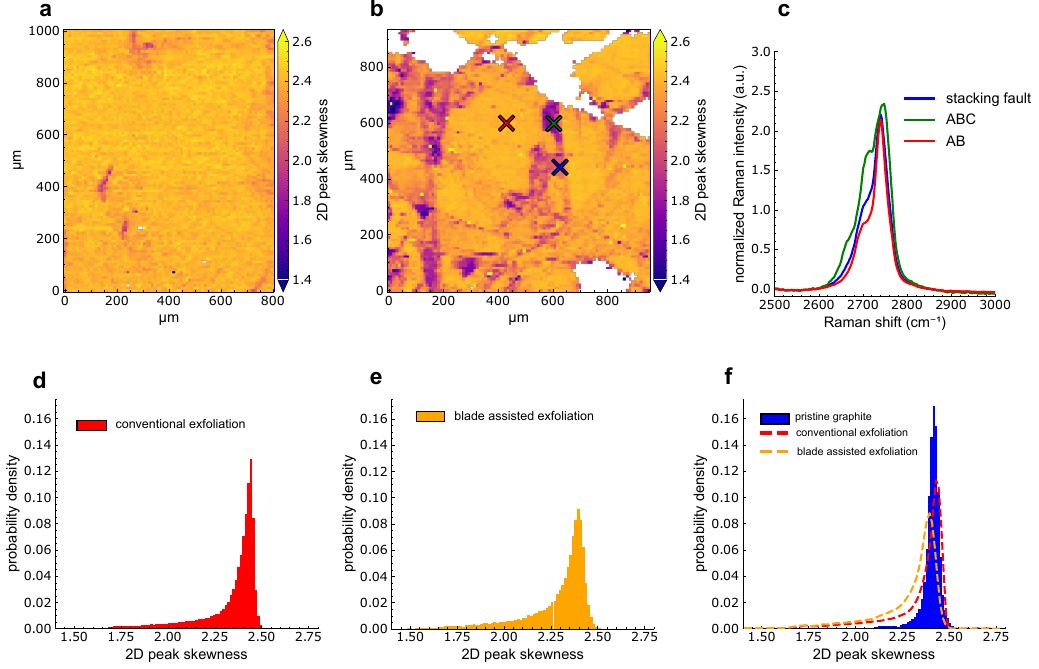}
    \caption{\textbf{Rhombohedral phase formation during the exfoliation step.}
    \textbf{a} Representative 2D peak skewness map of pristine natural graphite and \textbf{b} of blade-assisted exfoliated graphite, showing an increased abundance of low-skewness regions.
    \textbf{c} Representative Raman 2D bands extracted from AB, ABC-like, and stacking-fault regions marked by colored crosses in panel b.
    \textbf{d-f} Comparison of the probability density distributions of the 2D skewness for conventionally exfoliated, blade-assisted exfoliated thick graphite samples, and pristine graphite.}
    \label{fig:statistic_example}
\end{figure}

After defining the skewness-based classification, we applied this analysis to large-area Raman maps to quantify how different mechanical preparation routes influence the abundance of ABC-stacked regions.
We compared the statistical distribution of the 2D skewness for pristine natural graphite, conventionally exfoliated graphite, and blade-assisted exfoliated graphite.
Using 488 nm excitation, each preparation was sampled by multiple 1 $\upmu$m-resolution Raman maps summing to more than 3 mm$^2$ per sample type, equivalent to more than a thousand 50$\times$50 $\upmu$m$^2$ flakes.
We define the ABC area fraction as $F_{\mathrm{ABC}} = N_{\mathrm{ABC}} / N_{\mathrm{graphite}}$, where $N_{\mathrm{ABC}}$ is the number of pixels classified as ABC-like and $N_{\mathrm{graphite}}$ is the number of pixels covered by graphite.
The skewness window $1.6 < \gamma < 2.25$ admits any 2D line shape with reduced asymmetry, including perfect rhombohedral stacking, mixed AB-ABC stacks, and stacking-fault regions (Fig.~\ref{fig:statistic_example}c).
$F_{\mathrm{ABC}}$ is therefore an upper bound on the perfect-rhombohedral area fraction, with the exact partition between pure ABC and mixed/fault contributions requiring ERS~\cite{Palinkas2024-sm}.
At 488 nm excitation, the laser penetration depth in graphite is approximately 15 nm~\cite{Klar2013-rg}.
In the pristine source crystal this makes $F_{\mathrm{ABC}}$ a near-surface metric.
The exfoliated thick flakes considered in the statistical analysis are thinner than the penetration depth, as confirmed by the appearance of the Si substrate peak in the spectra, so the corresponding $F_{\mathrm{ABC}}$ values reflect the full thickness of the flake.
For the pristine graphite source crystal, the near-surface ABC area fraction is 3.1\% (Table~\ref{tab:abc_yield}), exceeding the volumetric value of $< 1\%$ measured by XRD on the same crystal (see Supplementary section S5).
The discrepancy is consistent with the surface-only nature of the Raman probe on the bulk crystal.
The tape-based cleaning step may also contribute a modest excess of ABC stacking near the surface.

\begin{table}[!htbp]
\centering
\small
\setlength{\tabcolsep}{5pt}
\renewcommand{\arraystretch}{1.15}
\begin{tabular}{|l|c|c|c|c|}
\hline
\multirow{2}{*}{\begin{tabular}{@{}l@{}}\textbf{exfoliation method} \\ \textbf{/ sample type}\end{tabular}} 
& \multirow{2}{*}{\begin{tabular}{@{}c@{}}\textbf{pristine} \\ \textbf{graphite}\end{tabular}} 
& \multicolumn{2}{c|}{\textbf{conventional exfoliation}} 
& \multirow{2}{*}{\begin{tabular}{@{}c@{}}\textbf{blade-assisted} \\ \textbf{exfoliation}\end{tabular}} \\ 
\cline{3-4}
&  & \textbf{ACH} & \textbf{Z-Z} &  \\ 
\hline
\textbf{total area (mm$^2$)} 
& 3.1 & 3.42 & 3.61 & 3.51 \\ 
\hline
\textbf{ABC area (mm$^2$)} 
& 0.1 & 0.54 & 0.6 & 0.92 \\ 
\hline
\textbf{area fraction, $F_{\mathrm{ABC}}$ (\%)} 
& 3.17 & 15.64 & 16.52 & 26.14 \\ 
\hline
\end{tabular}
\caption{\textbf{ABC area fractions for different exfoliation methods.}
The columns ACH and Z-Z denote conventional exfoliation with the tape-bending axis aligned, respectively, with the armchair and zigzag directions of the source crystal.
The numbers refer to thick graphite flakes (more than 20 layers), classified pixel-by-pixel from large-area Raman 2D-skewness maps at 488 nm excitation.
$F_{\mathrm{ABC}}$ is the fraction of graphite-covered pixels with $1.6 < \gamma < 2.25$.
Full classification windows and acquisition parameters are given in the Supporting Information, Sections S2 and S6.3.}
\label{tab:abc_yield}
\end{table}

We next report the ABC area fraction for standard (no-blade) exfoliation.
At the single-flake level, only an armchair-directed shear can convert AB into ABC, while zigzag-directed shear leaves the stacking unchanged~\cite{Nery2020-fe,Yang2019-er}.
To test whether this directional selectivity survives the macroscopic exfoliation process, we prepared two conventional-exfoliation series with the tape-bending direction aligned, respectively, with the armchair and zigzag axes of the source crystal.
The orientation of the source crystal was determined by scanning tunneling microscopy (see Supporting Information, Section S4, Figure S3).
The resulting ABC area fractions are 15.6\% and 16.5\% over more than 3 mm$^2$ of mapped surface each (Table~\ref{tab:abc_yield}).
The two values are comparable, with no statistically meaningful directional dependence at this length scale.
We interpret this as evidence that the macroscopic tape-bending direction does not impose a single shear orientation on every flake.
Each flake samples its own local bending and contact geometry on the tape and substrate.
The ABC yield resulting from exfoliation is therefore insensitive to the tape-bending direction.
Per-flake directional control is achievable in the case of single-flake transfer~\cite{Yang2019-er}, but better control of the shear direction is needed to achieve this by bulk exfoliation.
For blade-assisted exfoliation, the thick-flake ABC area fraction reaches 26.1\% (Table~\ref{tab:abc_yield}).
Representative skewness maps of the pristine graphite and of a thick flake prepared by blade-assisted exfoliation are shown in Fig.~\ref{fig:statistic_example}a,b, and the probability density distributions of the 2D peak skewness for all spectra of the source graphite surface, conventional exfoliation, and blade-assisted exfoliation are compared in Fig.~\ref{fig:statistic_example}d-f.
The skewness distribution broadens and develops a low-skewness tail with increasingly aggressive mechanical processing, in line with the area fractions in Table~\ref{tab:abc_yield}.

For thin flakes (layer number $< 20$), which are the practically relevant starting material for device work, the per-flake ABC area fraction reaches roughly 75\% in the upper tail of the distribution in the inset of Fig.~\ref{fig:flg_skew}a.
These results show that the rhombohedral content of the samples stems from the exfoliation procedure, modifications to which can result in a larger ABC area fraction~\cite{Holleis2026-qu}, and that increasing the shear and bending stresses during blade-assisted exfoliation can increase the fraction of ABC domains on individual FLG flakes, to a point where the preparation of samples is no longer bottlenecked by sample scarcity.
Almost any FLG flake one measures contains ABC domains.
Near-infrared optical microscopy~\cite{Feng2025-ob} can further speed up sample preparation by directly imaging the shape and extent of ABC domains, followed by exact identification of the stacking sequence via ERS~\cite{Palinkas2024-sm}.

\begin{figure}[htbp]
    \centering
    \includegraphics[width=0.5 \textwidth]{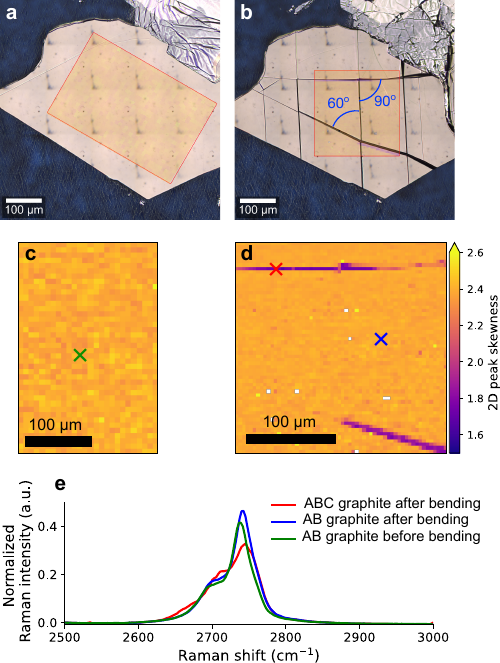}
    \caption{\textbf{Local formation of ABC domains induced by blade-assisted exfoliation.}
    Optical microscope image of the investigated thick graphite flake before \textbf{a} and after \textbf{b} blade-assisted exfoliation.
    The flake is supported on the exfoliation tape surface.
    The area selected for Raman mapping is indicated by red rectangles.
    Wrinkles formed during blade-assisted exfoliation meet mostly at angles that are multiples of 30$^{\circ}$.
    \textbf{c, d} Raman maps of the 2D peak skewness recorded before and after blade-assisted exfoliation, showing the local emergence of ABC-related contrast after mechanical deformation.
    \textbf{e} Representative Raman 2D bands extracted from the locations marked in the maps.}
    \label{fig:domain_formation}
\end{figure}

Next, we examine the perturbations caused by blade-assisted exfoliation on individual flakes, to highlight the changes that lead to rhombohedral graphite formation during exfoliation.
Figure~\ref{fig:domain_formation} shows the same graphite flake on the adhesive tape, before (a, c) and after (b, d) the blade treatment.
The blade-assisted exfoliation generates wrinkles that meet at angles that are multiples of 30$^{\circ}$, with 60$^{\circ}$ and 90$^{\circ}$ visible in Fig.~\ref{fig:domain_formation}b.
Raman measurements on these wrinkles indicate the presence of non-hexagonal stacking, consistent with the formation of AB-ABC domain walls in the wrinkles.
ABC-like spectra appear only on one of the two wrinkle sets and not on the perpendicular set, in line with the observation of Rooney et al.~\cite{Rooney2018-px} that the AB-ABC transition proceeds only along the armchair direction~\cite{Nery2020-fe,Yang2019-er}.
Wrinkle formation is therefore one direct source of AB-ABC domain walls, whose subsequent motion can sweep ABC stacking into otherwise purely hexagonal flakes.
A second flake, presented in the Supporting Information (Section S3, Figure S2), shows an example of wrinkle-ABC domain correlation: wrinkles delineate the boundaries of ABC domains.
This behavior is observed on most flakes we have investigated, with the ABC domains consistently starting at wrinkles.
The examples in Fig.~\ref{fig:domain_formation} and the Supporting Information are representative of this trend.

\begin{figure}[htbp]
    \centering
    \includegraphics[width=\textwidth]{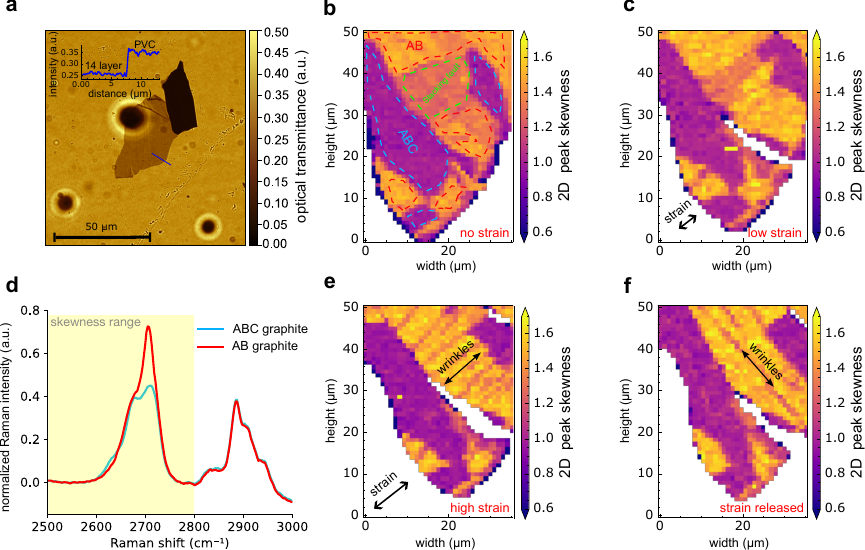}
    \caption{\textbf{Strain-induced evolution of stacking domains in graphite on a flexible substrate.}
    \textbf{a,} Optical transmission image of the graphite flake exfoliated onto PVC.
    The optical contrast was used to measure the thickness of the flake, corresponding to 14 graphene layers.
    \textbf{b,c,e,f,} Raman 2D-skewness maps of the same flake measured under different mechanical conditions: no strain (\textbf{b}), low strain (\textbf{c}), high strain (\textbf{e}), and after strain release (\textbf{f}).
    The maps reveal substantial evolution of the domain structure upon stretching, including domain displacement, the appearance of new domain boundaries, and the rearrangement of stacking-fault regions.
    For graphite on PVC, the 2D skewness was computed over a truncated $2500$--$2800$~cm$^{-1}$ window (shaded in panel \textbf{d}) to exclude the large PVC Raman peak near 2900~cm$^{-1}$ (see Supporting Information, Section S1).
    Smaller PVC features remaining within this window further shift the calculated skewness, producing the color-scale offset relative to Figs.~\ref{fig:flg_skew}--\ref{fig:domain_formation}.
    Low skewness within each map still marks ABC-rich regions.
    \textbf{d,} Representative Raman 2D bands extracted from the mapped area, showing the characteristic spectral signatures of AB and ABC graphite.
    The shaded region indicates the skewness range used for domain analysis.}
    \label{fig:strain}
\end{figure}

The bending and shearing of the exfoliation tape, even during standard exfoliation, can induce local stretching and compression of the graphite flakes.
To investigate the influence of this mechanical strain on the domain-wall dynamics, we performed controlled deformation of individual FLG flakes on a polyvinyl chloride (PVC) membrane mounted in a custom stretching device (Supporting Information, Figure S5), and applied uniaxial strain while measuring Raman maps on selected flakes.
One example is shown in Fig.~\ref{fig:strain}, where we apply and then relax uniaxial strain on a 14-layer FLG flake, modeling the strain experienced during exfoliation.
Comparing panels b and c of Fig.~\ref{fig:strain}, we observe that at the onset of uniaxial strain the domains start to rearrange, indicating domain-wall movement triggered by mechanical deformation.
One region, identified as imperfect stacking in panel b (green outline), has vanished, indicating that layer configurations with high stacking energy transform into lower-energy ones~\cite{Roy2025-su}.
Upon applying higher strain, further movement of the domain walls can be observed, as well as wrinkles appearing parallel to the strain direction (see Fig.~\ref{fig:strain}e).
We attribute this to the difference in Poisson's ratio between graphite (0.16~\cite{Blakslee1970-pd} to 0.19~\cite{Politano2015-xa}) and PVC (0.35 to 0.42~\cite{Faccinto2025-gn}).
This difference leads to a compression perpendicular to the applied strain direction.
On strain release (Fig.~\ref{fig:strain}f), wrinkles appear perpendicular to the previously applied strain direction, likely from slippage of the flake on the PVC that leaves the flake in residual compression.
The wrinkles themselves show lower 2D skewness, confirming the formation of further AB-ABC domain walls.
Another example of strain-induced domain rearrangement is shown in the Supporting Information (Section S6, Figure S6), where the domain walls move under uniaxial strain, and even an AB-stacked domain transforms into ABC.

Conventional exfoliation and blade-assisted exfoliation both act through the same physical channel: an interlayer shear, driven by bending of the flake, that nucleates AB-ABC domain walls along the armchair direction.
Mechanical strain experienced by the flakes on the tape and during exfoliation can subsequently move the domain walls across the flake.
Each mechanical perturbation therefore contributes both to the creation of new rhombohedral domains and to the rearrangement of an existing stacking landscape.

Mechanical exfoliation has produced the cleanest 2D-material samples for two decades.
The present result shows that it has been doing more than isolating few-layer flakes.
For blade-assisted exfoliation of thin flakes, the ABC area fraction rises significantly, reaching 75\% in the upper tail of the distribution.
Direct imaging shows that the underlying process is the nucleation and motion of AB-ABC domain walls driven by interlayer shear, and externally applied strain can reorganize the stacking landscape on a single-flake level.
Enhanced bending and shearing during exfoliation give a simple preparation method for ABC-rich graphite, supporting systematic studies of the correlated and topological phases of the rhombohedral phase~\cite{Han2023-eb,Hagymasi2022-hg,Liu2023-gh,Zhou2021-of,Yang2025-ie}.
The AB-ABC domain walls we image are analogs of the solitons and partial dislocations studied in bilayer graphene~\cite{Alden2013-ol}, of the stacking boundaries that underlie sliding ferroelectricity in bilayer h-BN~\cite{Yasuda2024-wh}, and of the reconstructed stacking domains found in marginally twisted moir\'e superlattices~\cite{Halbertal2021-lq}.
Blade-assisted exfoliation could therefore provide a method to introduce domain walls in these systems.

\section*{Experimental methods}
Samples were exfoliated using ``blue tape'' (Ultron Systems, P/N 1008R-8.0).
As substrate we used Si wafers with 90~nm SiO$_2$.
Natural graphite samples were purchased from NGS Trading \& Consulting GmbH (www.graphit.de).
All flakes presented in this paper were exfoliated from the same side of a large graphite crystal (shown in Supporting Information, Figure S3).
XRD measurements were performed on this crystal as well.
Raman measurements were performed on a WITec 300rsa+ confocal Raman system, using 488~nm laser excitation.
Data processing and figure generation were performed using the open-source Python tool Ramantools~\cite{Nemes-Incze2023-ea}.
For more details, see Section S6 of the Supporting Information.

\section*{Data availability}
The raw data required to reproduce these findings are deposited in Zenodo at [DOI Link].

\section*{Author Contributions}
KM prepared the samples and performed the Raman characterization and data analysis.
KK performed STM measurements.
GK assisted in the measurements on a PVC support.
ZsEH performed XRD measurements and data analysis.
KND and GyK assisted in sample preparation.
LT contributed to data analysis and interpretation.
KM and PNI wrote the manuscript with contributions from all authors.
PNI conceived and coordinated the project.

\section*{Acknowledgments}
The work was conducted within the framework of the MTA - HUN-REN EK Lendület ``Momentum'' Topology in Nanomaterials Research Group through project LP2024-17.
Financial support from NKFIH through grants \'{E}lvonal KKP 138144, Excellence 151372, K146156, and TKP2021-NKTA-05 is also acknowledged.

\newpage

\bibliography{RG_prep_refs}

\end{document}